\documentclass[prl,aps,twocolumn,preprintnumbers,showpacs]{revtex4}
\usepackage{epsfig,amssymb,amsfonts}
\def\id{\mathbb{I}}
\newcommand{\ket}[1]{\left | \, #1 \right\rangle}
\newcommand{\bra}[1]{\left \langle #1 \, \right |}
\newcommand{\proj}[1]{\ket{#1}\bra{#1}}
\newcommand{\braket}[2]{\left\langle\, #1\,|\,#2\,\right\rangle}
\newcommand{\outprod}[2]{\ket{#1}\bra{#2}}

\def\opone{\leavevmode\hbox{\small1\kern-3.8pt\normalsize1}}
\newcommand{\tr}[1]{\mbox{Tr} \, #1 }
\newcommand{\vis}{\text{v}}

\begin{document}
\title{Direct estimations of linear and non-linear functionals of a quantum state}
\date{\today}
    \author{Artur K. \surname{Ekert}}
    \author{Carolina \surname{Moura Alves}}\email{carolina.mouraalves@qubit.org}%
    \author{Daniel K. L. \surname{Oi}}%
    \affiliation{Centre for Quantum Computation, Clarendon Laboratory,
      University of Oxford, Parks Road, Oxford OX1 3PU, U.K.}%
    \author{Micha{\l} \surname{Horodecki}}%
    \affiliation{ Institute of Theoretical Physics and Astrophysics,
      University of Gda\'nsk, 80-952 Gda\'nsk, Poland.}
    \author{Pawe{\l} \surname{Horodecki}}%
    \affiliation{Faculty of Applied Physics and Mathematics, Technical
      University of Gda\'nsk, 80-952
      Gda\'nsk, Poland.} %
    \author{L. C. \surname{Kwek}}%
    \affiliation{Department of Natural Sciences, National Institute of
      Education, Nanyang Technological University, 1 Nanyang Walk, Singapore
      637616}

\begin{abstract}
  We present a simple quantum network, based on the controlled-SWAP
  gate, that can extract certain properties of quantum states
  without recourse to quantum tomography. It can be used used as a
  basic building block for direct quantum estimations of both
  linear and non-linear functionals of any density operator. The
  network has many potential applications ranging from purity
  tests and eigenvalue estimations to direct characterization of
  some properties of quantum channels. Experimental realizations
  of the proposed network are within the reach of quantum
  technology that is currently being developed.
\end{abstract}


\maketitle

\preprint{Version 10}

Certain properties of a quantum state $\varrho$, such as its
purity, degree of entanglement, or its spectrum, are of
significant importance in quantum information science. They can be
quantified in terms of linear or non-linear functionals of
$\varrho$. Linear functionals, such as average values of
observables $\{A\}$, given by $\tr A\varrho$, are quite common as
they correspond to directly measurable quantities. Non-linear
functionals of state, such as the von Neumann entropy
$-\tr\varrho\ln\varrho$, eigenvalues, or a measure of purity
$\tr\varrho^2$, are usually extracted from $\varrho$ by classical
means i.e. $\varrho$ is first estimated and once a sufficiently
precise classical description of $\varrho$ is available, classical
evaluations of the required functionals can be made. However, if
only a limited supply of physical objects in state $\varrho$ is
available, then a direct estimation of a specific quantity may be
both more efficient and more desirable~\cite{direct}. For example,
the estimation of purity of $\varrho$ does not require knowledge
of all matrix elements of $\varrho$, thus any prior state
estimation procedure followed by classical calculations is, in
this case, inefficient. However, in order to bypass tomography and
to estimate non-linear functionals of $\varrho$ more directly, we
need quantum networks~\cite{Deu89,BBC95+} performing quantum
computations that supersede classical evaluations.

In this paper, we present a simple quantum network which can be
used as a basic building block for direct quantum estimations of
both linear and non-linear functionals of any $\varrho$. The
network can be realized as multiparticle interferometry. While
conventional quantum measurements, represented as quantum networks
or otherwise, allow the estimation of $\tr A\varrho$ for some Hermitian
$A$, our network can also provide a direct estimation of the
overlap of any two unknown quantum states $\varrho_a$ and
$\varrho_b$, i.e. $\tr\varrho_a\varrho_b$. Here, and in the
following, we use terminology developed in quantum information
science. For a comprehensive overview of this terminology,
including quantum logic gates and quantum networks see, for
example,~\cite{NielsenChuang}.

\begin{figure}[tbp]
  \epsfig{figure=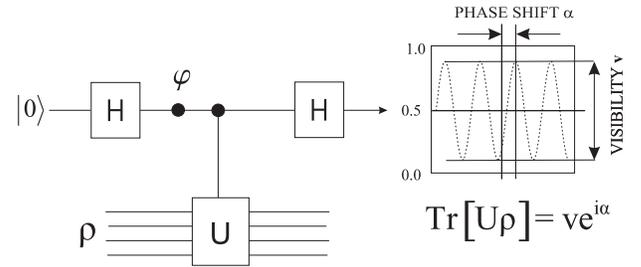,width=0.45\textwidth} \caption{Both the
    visibility and the shift of the interference patterns of a single qubit
    (top line) are affected by the controlled-$U$ operation on a general
    state, $\rho$.} \label{figint}
\end{figure}

In order to explain how the network works, let us start with a
general observation related to modifications of visibility in
interferometry. Consider a typical interferometric set-up for a
single qubit: Hadamard gate, phase shift $\varphi$, Hadamard gate,
followed by a measurement in the computational basis. We modify
the interferometer by inserting a controlled-$U$ operation between
the Hadamard gates, with its control on the qubit and with $U$
acting on a quantum system described by some unknown density
operator $\rho$, as shown in Fig.~\ref{figint}. The action of the
controlled-$U$ on $\rho$ modifies the interference pattern by the
factor~\cite{SPEAEOV2000},
\begin{equation}
\tr\rho U = \vis e^{i\alpha},
\label{eqvisi}
\end{equation}
where $\vis$ is the new visibility and $\alpha$ is the shift of
the interference fringes, also known as the Pancharatnam
phase~\cite{Pancha56}. The observed modification of the visibility
gives an estimate of $\tr U\rho$, i.e. the average value of the
unitary operator $U$ in state $\rho$. Let us mention in passing
that this property, among other things, allows to estimate an
unknown quantum state $\rho$ as long as we can estimate $\tr
U_k\rho$ for a set of unitary operators $U_k$ which form a basis
in the vector space of density operators.

\begin{figure}[tbp]
\epsfig{figure=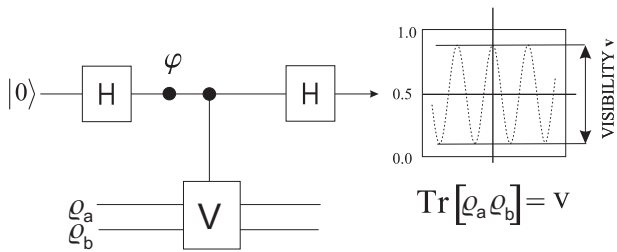,width=0.45\textwidth} \caption{A
quantum network for direct estimations of both linear and
non-linear functions of state. The probability of finding the
control (top line) qubit in state $\ket{0}$ at the output depends
on the overlap of the two target states (two bottom lines). Thus
estimation of this probability leads directly to an estimation of
$\tr \varrho_a\varrho_b=\vis=2\,P_0-1$.} \label{figdev}
\end{figure}

Let us now consider a quantum state $\rho$ of two separable subsystems, such
that $\rho =\varrho_{a}\otimes\varrho_{b}$. We choose our controlled-$U$ to be
the controlled-$V$, where $V$ is the swap operator, defined as,
$V\ket{\alpha}_{A}\ket{\beta}_{B}=\ket{\beta}_{A}\ket{\alpha}_{B}$, for any
pure states $\ket{\alpha}_{A}$ and $\ket{\beta}_{B}$. In this case, the modification
of the interference pattern given by Eq.~(\ref{eqvisi}) can be written as,
\begin{equation}
\vis=\tr V\left(\varrho _{a}\otimes\varrho _{b}\right)=\tr\varrho_{a}\varrho _{b}.
\end{equation}
which is easily proved using the spectral decomposition of $\varrho_a$ and
$\varrho_b$. Since $\tr \varrho_{a}\varrho _{b}$ is real, we can
fix $\varphi=0$ and the probability of finding the qubit in state
$\ket{0}$ at the output, $P_0$, is related to the visibility by $
\vis=2\,P_0-1$. This construction, shown in Fig.~(\ref{figdev}),
provides a direct way to measure $\tr\varrho_{a}\varrho _{b}$
(c.f.~\cite{Filip2001} for a related idea).

There are many possible ways of utilizing this result. For pure
states $\varrho_a=\proj{\alpha}$ and $\varrho_b=\proj{\beta}$ the
formula above gives $\tr\varrho_a\varrho_b
=|\braket{\alpha}{\beta}|^2$ i.e. a direct measure of
orthogonality of $\ket{\alpha}$ and $\ket{\beta}$. If we put
$\varrho_a=\varrho_b=\varrho$ then we obtain an estimation of
$\tr\varrho^2$. In the single qubit case, this measurement allows
us to estimate the length of the Bloch vector, leaving its
direction completely undetermined. For qubits $\tr\varrho^2$ gives
the sum of squares of the two eigenvalues which allows to estimate
the spectrum of $\varrho$.

In general, in order to evaluate the spectrum of any $d \times d$
density matrix $\varrho$ we need to estimate $d-1$ parameters
${\rm Tr}\varrho^2$, ${\rm Tr}\varrho^3$,... ${\rm Tr}\varrho^d$.
For this we need the controlled-shift operation, which is a
generalization of the controlled-swap gate. Given $k$ systems of
dimension $d$ we define the shift $V^{(k)}$ as
\begin{equation}
V^{(k)} \ket{\phi_1}\ket{\phi_2}...\ket{\phi_k} =
\ket{\phi_k}\ket{\phi_1}...\ket{\phi_{k-1}},
\end{equation}
for any pure states $\ket{\phi}$. Such an operation can be easily
constructed by cascading $k-1$ swaps $V$. If we extend the network
and prepare $\rho=\varrho^{\otimes k}$ at the input then the
interference will be modified by the visibility factor,
\begin{equation}
\vis = {\rm Tr}\, V^{(k)}\varrho^{\otimes k} = {\rm Tr}\,
\varrho^k = \sum_{i=1}^m {\lambda_i}^k. \label{powers}
\end{equation}
Thus measuring the average values of $V^{(k)}$ for $k=2,3...d$ allows us to
evaluate the spectrum of $\varrho$~\cite{direct}.  Although we have not
eliminated classical evaluations, we have reduced them by a significant
amount. The average values of $V^{(k)}$ for $k=2,3...d$ provide just enough
information to evaluate the spectrum of $\varrho$ but certainly not enough to
estimate the whole density matrix.

So far we have treated the two inputs, $\varrho_a$ and $\varrho_b$
in a symmetric way. However, there are several interesting
applications in which one of the inputs, say $\varrho_a$, is
predetermined and the other is unknown. For example, projections
on a prescribed vector $\ket{\psi}$, or on the subspace
perpendicular to it, can be implemented by choosing
$\varrho_a=\proj{\psi}$. By changing the input state $\ket{\psi}$
we effectively ``reprogram'' the action of the network which then
performs different projections. This property can be used in
quantum communication, in a scenario where one carrier of
information, in state $\ket{\psi}$, determines the type of
detection measurement performed on the second carrier. N.B. as the
state $\ket{\psi}$ of a single carrier cannot be determined, the
information about the type of the measurement to be performed by
the detector remains secret until the moment of detection.


Another interesting application is the estimation of the extremal eigenvalues
and eigenvectors of $\varrho_{b}$ without reconstructing the entire spectrum.
In this case, the input states are of the form $\proj{\psi}\otimes\varrho_{b}$
and we vary $\ket{\psi}$ searching for the minimum and the maximum of
$\vis=\bra{\psi}\varrho_{b}\ket{\psi}$.  This, at first sight, seems to be a
complicated task as it involves scanning $2(d-1)$ parameters of $\psi$.  The
visibility is related to the overlap of the reference state $\ket{\psi}$ and
$\varrho_{b}$ by,
\begin{eqnarray}
\vis_{\psi}&=&\tr\left(\proj{\psi}\sum_{i}\lambda_{i}\proj{\eta_{i}}\right)\nonumber\\
 &=&\sum_{i}\lambda_{i}\left|\braket{\psi}{\eta_{i}}\right|^2=\sum_{i}\lambda_{i}p_{i},
\end{eqnarray}
where $\sum_{i}p_{i}=1$. This is a convex sum of the eigenvalues of
$\varrho_{b}$ and is minimized (maximized) when
$\ket{\psi}=\ket{\eta_{min}}\;\left(\ket{\eta_{max}}\right)$. For any
$\ket{\psi}\neq\ket{\eta_{min}}\;\left(\ket{\eta_{max}}\right)$, there exists
a state, $\ket{\psi'}$, in the neighbourhood of $\ket{\psi}$ such that
$\vis_{\psi'}<\vis_{\psi}$ ($\vis_{\psi'}>\vis_{\psi}$), thus this global
optimization problem can easily be solved using standard iterative methods,
such as steepest decent~\cite{Gill1981}.

Estimation of extremal eigenvalues plays a significant role in the
direct detection~\cite{direct} and
distillation~\cite{Horodeckis1997} of quantum entanglement. For
example, in a special case if two qubits described by the density
operator $\varrho_{b}$, such that the reduced density operator of
one of the qubits is maximally mixed, we can test for the
separability of $\varrho_{b}$ by checking whether the maximal
eigenvalue of $\varrho_{b}$ does not exceed $\frac{1}{2}$~\cite{xor}.

Finally, we may want to estimate an unknown state, say a $d\times
d$ density operator, $\varrho_{b}$. Such an operator is determined
by $d^2-1$ real parameters. In order to estimate matrix elements
$\bra{\psi}\varrho_{b}\ket{\psi}$, we run the network as many
times as possible (limited by the number of copies of
$\varrho_{b}$ at our disposal) on the input
$\proj{\psi}\otimes\varrho_{b}$, where $\ket{\psi}$ is a pure
state of our choice. For a fixed $\ket{\psi}$, after several runs,
we obtain an estimation of,
\begin{equation}
\vis=\bra{\psi}\varrho_{b}\ket{\psi}.
\end{equation}
In some chosen basis $\{\ket{n}\}$ the diagonal elements
$\bra{n}\varrho_{b}\ket{n}$ can be determined using the input
states $\proj{n}\otimes\varrho_{b}$. The real part of the
off-diagonal element $\bra{n}\varrho_{b}\ket{k}$ can be estimated
by choosing $\ket{\psi}=(\ket{n}+\ket{k})/\sqrt{2}$, and the
imaginary part by choosing $\ket{\psi}=(\ket{n}+i
\ket{k})/\sqrt{2}$. In particular, if we want to estimate the
density operator of a qubit, we can choose the pure states,
$\ket{0}$ (spin +$z$), $\left(\ket{0}+\ket{1}\right)/\sqrt{2}$
(spin +$x$) and $\left(\ket{0}+i \ket{1}\right)/\sqrt{2}$ (spin
+$y$), i.e. the three components of the Bloch vector.

Needless to say, quantum tomography can be performed in many other
ways, the practicalities of which depend on technologies involved.
However, it is worth stressing that our scheme is based on a
network of a fixed architecture which is controlled only by input
data, a feature that can be useful in some quantum communication
scenarios.

We can extend the procedure above to cover estimations of
expectation values of arbitrary observables. This can be done with
the network shown in Fig.~\ref{figdev} because estimations of mean
values of \emph{any} observable  can always be reduced to
estimations of a binary two-output POVMs~\cite{PH}.We shall apply
the technique developed in Refs.~\cite{SPA,direct}. As
$A'=\gamma\id+A$ is positive if $-\gamma$ is the minimum negative
eigenvalue of $A$, we can construct the state
$\varrho_{a}=\varrho_{A'}=\frac{A'}{\tr(A')}$ and apply our
interference scheme to the pair $\varrho_{A'}\otimes\varrho_{b}$.
The visibility gives us the mean value of V,
\begin{equation}
\vis=\langle V\rangle_{\varrho_{A'}\otimes\varrho_{b}}
=\tr\left(\frac{A'}{\tr(A')}\varrho_{b}\right),
\end{equation}
which leads us to the desired value,
\begin{equation}
\langle A\rangle_{\varrho_{b}}\equiv\tr(\varrho_{b} A)=\vis\tr A+\gamma(\vis d-1),
\end{equation}
where $\tr\id=d$.

\begin{center}
\begin{figure}
  \epsfig{figure=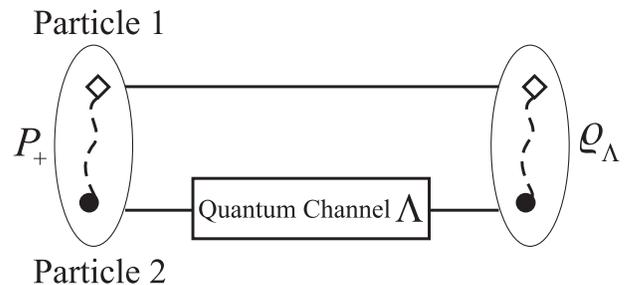,width=0.45\textwidth}
  \caption{A quantum channel $\Lambda$ acting on one of the subsystems of a bipartite maximally
    entangled state of the form $\ket{\psi_+}= \sum_k\ket{k}\ket{k}/\sqrt{d}$.
    $P_+$ is the corresponding density operator, i.e. $P_+= \proj{\psi_+}$.
    The output is the state described by the density operator
    $\varrho_{\Lambda}=
    \frac{1}{d}\sum_{kl}\outprod{k}{l}\otimes\Lambda\left(\outprod{k}{l}\right)$,
    which contains a complete information about the channel. This isomorphism
    between $\Lambda$ and $\varrho_{\Lambda}$ allows to infer all properties
    of the channel from the corresponding properties of the state. Any
    subsequent estimations of $\varrho_{\Lambda}$, or any of its functions,
    provides information about the completely positive map $\Lambda$.}
  \label{figchan}
\end{figure}
\end{center}

Any technique that allows direct estimations of properties of
quantum states can be also used to estimate certain properties of
quantum channels. Recall that, from a mathematical point of view,
a quantum channel is a trace preserving linear map,
$\varrho\rightarrow\Lambda(\varrho)$, which maps density operators
into density operators, and whose trivial extensions,
$\mathcal{I}_{k}\otimes\Lambda$ do the same, i.e. $\Lambda$ is a
completely positive map. In a chosen basis the action of the
channel on a density operator
$\varrho=\sum_{kl}\varrho_{kl}\outprod{k}{l}$ can be written as
\begin{equation}
\Lambda (\varrho) = \Lambda
\left(\sum_{kl}\varrho_{kl}\outprod{k}{l}\right) =
\sum_{kl}\varrho_{kl}\Lambda\left(\outprod{k}{l}\right).
\end{equation}
Thus the channel is completely characterized by operators
$\Lambda\left(\outprod{k}{l}\right)$. In fact, with every channel
$\Lambda$  we can associate a quantum state $\varrho_{\Lambda}$
which provides a complete characterization of the channel. For if
we prepare a maximally entangled states of two particles described
by the density operator
$P_{+}=\frac{1}{d}\sum_{kl}\outprod{k}{l}\otimes\outprod{k}{l}$,
and if we send only one particle through the channel, as shown in
Fig.~\ref{figchan}, then we obtain
\begin{equation}
\label{varrholambda}
P_{+}\rightarrow\left[\mathcal{I}\otimes\Lambda\right]
P_{+}=\varrho_{\Lambda},
\end{equation}
where
\begin{equation}
\varrho_{\Lambda}=\frac{1}{d}\sum_{kl}\outprod{k}{l}\otimes\Lambda\left(\outprod{k}{l}\right).
\end{equation}

We may interpret this as mapping the $\outprod{k}{l}^{th}$-element
of an input density matrix to the output matrix,
$\Lambda\left(\outprod{k}{l}\right)$. Thus, knowledge of
$\varrho_{\Lambda}$ allows us to determine the action of $\Lambda$
on an arbitrary state, $\varrho\rightarrow\Lambda(\varrho)$. If we
perform a state tomography on $\varrho_{\Lambda}$ we effectively
perform a quantum channel tomography. If we choose to estimate
directly some functions of $\varrho_{\Lambda}$ then we gain some
knowledge about specific properties of the channel without
performing the full tomography of the channel.

For example, consider a single qubit channel. Suppose we are
interested in the maximal rate of a reliable transmission of
qubits per use of the channel, which can be quantified as the
channel capacity. However, unlike in the classical case, quantum
channels admit several capacities~\cite{Horodeckis2000,huge},
because users of quantum channels can also exchange classical
information. We have then the capacities $Q_{C}$ where
$C=\o,\leftarrow, \rightarrow, \leftrightarrow$, stands for zero
way, one way and two way classical communication. In general, it
is very difficult to calculate the capacity of a given channel.
However, our extremal eigenvalue estimation scheme provides a
simple necessary and sufficient condition for a one qubit channel
to have non-zero two-way capacity. Namely, $Q_{\leftrightarrow}>0$
iff $\varrho_{\Lambda}$ has maximal eigenvalue greater than
$\frac{1}{2}$. (Clearly, this  a necessary condition for the other
three capacities to be non-zero).

This result becomes apparent by noticing that if we trace
$\varrho_{\Lambda}$ over the qubit that went through the channel
$\Lambda$ (particle 2 in Fig.~\ref{figchan}), we obtain the
maximally mixed state. Furthermore, the two qubit state,
$\varrho_{\Lambda}$, is two-way distillable iff the operator
$\frac{\id}{2} \otimes\id - \varrho_{\Lambda}$ has a negative
eigenvalue (see~\cite{xor} for details), or equivalently, when
$\varrho_{\Lambda}$ has the maximal eigenvalue greater than
$\frac{1}{2}$. This implies $Q_{\leftrightarrow}(\Lambda)>0$
because two-way distillable entanglement, which is non-zero iff
given state is two way distillable, is the lower bound for
$Q_{\leftrightarrow}(\Lambda)$~\cite{huge}.


In summary, we have described a simple quantum network which can
be used as a basic building block for direct quantum estimations
of both linear and non-linear functionals of any density operator
$\varrho$. It provides a direct estimation of the overlap of any
two unknown quantum states $\varrho_a$ and $\varrho_b$, i.e.
$\tr\varrho_a\varrho_b$. Its straightforward extension can be
employed to estimate functionals of any powers of density
operators. The network has many potential applications ranging
from purity tests and eigenvalue estimations to direct
characterization of some properties of quantum channels.

Finally let us also mention that the controlled-SWAP operation is
a direct generalization of a Fredkin gate~\cite{FT82} and can be
constructed out of simple quantum logic gates~\cite{BBC95+}. This
means that experimental realizations of the proposed network are
within the reach of quantum technology that is currently being
developed~(for an overview see, for example,~\cite{BEZ}).

A.K.E. and L.C.K. acknowledge financial support provided under the
NSTB Grant No. 012-104-0040. P.H. and M.H. would like to
acknowledge support from the Polish Committee for Scientific
Research and the European Commission. C.M.A. is supported by the
Funda{\c c}{\~a}o para a Ci{\^e}ncia e Tecnologia (Portugal) and
D.K.L.O would like to acknowledge the support of CESG (UK) and
QAIP (contract no. IST-1999-11234).


\end{document}